\def\etal{{\it et al.\/}\ }

\def\lsim{~\rlap{$<$}{\lower 1.0ex\hbox{$\sim$}}}
\def\gsim{~\rlap{$>$}{\lower 1.0ex\hbox{$\sim$}}}
\def\void#1{{}}

\documentclass[runningheads]{cl2emult}

\usepackage{makeidx}  
\usepackage{graphicx} 
\usepackage{subeqnar} 
\usepackage{multicol} 
\usepackage{cropmark} 
\makeindex            

\begin{document}
\title*{Optical/Infrared Survey of Galaxy Clusters}
%
%
%
%
\titlerunning{Optical/Infrared Survey of Galaxy Clusters}
%

\author{Luiz da Costa\inst{1}
\and S. Arnouts\inst{1}
\and S. Bardelli\inst{2}
\and C. Benoist\inst{3}
\and A. Biviano\inst{4}
\and S. Borgani\inst{5}
\and W. Boschin \inst{7}
\and T. Erben \inst{6}
\and M. Girardi \inst{7}
\and H. E. J{\o}rgensen  \inst{8}
\and L. F. Olsen  \inst{8}
\and M. Ramella \inst{4}
\and M. Schirmer \inst{6}
\and P. Schneider \inst{9}
\and M. Scodeggio \inst{10}
\and E. Zucca \inst{2}}

\authorrunning{da Costa \etal}
%
%
\institute{European Southern Observatory, Karl-Schwarzschild-Str 2,
 D-85728,  Garching bei M\"unchen,  Germany
\and
Osservatorio Astronomico di
Bologna, via Ranzani 1, I-40127, Bologna, Italy
\and 
Observatoire de la Cote d'Azur-Nice, BP4229, F-06304, Nice, cedex 4, France
\and
Osservatorio Astronomico di Trieste, via  Tiepolo 11, I-34131 Trieste, Italy
\and
INFN, Sezione di Trieste, Dipartimento
di Astronomia, Universit\'a degli Studi di Trieste, via Tiepolo 11,
I-34100, Trieste, Italy
\and
Max Planck Institut fur Astrophysiks, Karl-Schwarzschild-Str. 1,
Postfach 1523,
D-85740,  Garching bei M\"unchen, Germany
\and
Dipartimento di Astronomia, Universit\'a degli Studi di
                     Trieste, via  Tiepolo 11, I-34100, Trieste, Italy
\and
Copenhagen University Observatory,
Juliane Maries Vej 30, 2100, Copenhagen, Denmark
\and
 Institut f\"ur Astrophysik und Extraterrestrische Forschung 
der Universit\"at Bonn, Auf dem H\"ugel 71, D-53121 Bonn, Germany
\and
Istituto di Fisica Cosmica "G. Occhialini", via Bassini 15, I-20133,
Milano, Italy }

\maketitle              

\begin{abstract} In this contribution the ongoing effort to build  a
statistical sample of clusters of galaxies over a wide range of
redshifts to study the evolution of clusters and member galaxies is
reviewed. The starting point for this project has been the list of
candidate clusters identified from the $I$-band EIS-WIDE survey data.
Since the completion of this survey, new optical/infrared observations
have become available and have been used to confirm some of these
candidates using the photometric data alone or in combination with the
results of follow-up spectroscopic observations. Our preliminary
results show that the yield of real physical associations from the
original catalog is conservatively $\gsim60\%$ and that a large sample
of clusters in the southern hemisphere, extending to high-redshifts,
is within reach.
\end{abstract}

\section {Introduction}

Clusters of galaxies have always been objects of great interest for
studies of the origin and evolution of structures. They represent the
most massive virialized systems, being formed at the tail of the mass
distribution function. Their abundance is critically dependent on the
cosmological parameters and they offer a unique opportunity to study
the evolution of galaxies and the role of environmental
effects. Furthermore, there is currently unambiguous evidence for the
existence of massive clusters at much larger redshifts than originally
anticipated, thus providing a large time baseline for evolutionary
studies.

Unfortunately, current samples at high-redshifts ($z\gsim0.5$),
consisting primarily of X-ray selected systems, are still small and are
predominantly located in the northern hemisphere and thus inaccessible
to VLT.  The primary goals of the present effort are to: define a
statistical sample of clusters of galaxies over a broad range of
redshifts in the southern hemisphere; study their dynamics and measure
their mass; and to study the properties of the member galaxies. Carrying
out such a study as a function of the look-back time should provide both
constraints on cosmological parameters and greater insight into the
formation and evolution of structures in the universe.

The present work is based on a list of over 300 optically-selected
cluster candidates identified by applying the matched-filter algorithm
to the $I$-band data of the EIS-WIDE survey, covering 17 square
degrees of the southern
sky~\cite{olsen1},\cite{olsen2},\cite{marco}. While the deficiencies
of an optically-selected sample are well-known, there are also a
number of potential advantages. First, such samples are an important
complement to X-ray selected samples which are likely to select
preferentially massive, virialized systems.  Second, comparison
between X-ray and optical/infrared selected samples is of great
interest as it may provide information on the evolution of the
intra-cluster gas. Third, the sample of candidate clusters can grow in
size at a modest cost.

Though a necessary first step, having a sample of cluster candidates
identified in a single passband falls short of what is required for 
an efficient follow-up spectroscopic program. For that it is necessary to
have both an understanding of the survey yield, requiring the
confirmation of a significant number of candidates, and additional data
to enable the selection of candidate cluster galaxies. For this purpose,
we have for the past two years gathered multi-color data to obtain
supporting evidence for the reality of the cluster candidates, measure
the yield of real systems as well as to prune the sample of false
positives and identify candidate cluster members, based on color or
photometric redshifts, to serve as targets for spectroscopic
observations.

In this paper we present some preliminary results that have been
obtained from ongoing follow-up imaging (Section~2) and spectroscopic
observations (Section~3). Our future plans are summarized in
Section~4.

\section {Multi-color and Deep Imaging Data}

To provide additional leverage to the confirmation of the cluster
candidates extracted from the $I$-band survey, data taken in different
passbands have been combined. These data have been compiled from
publicly available EIS data or from pointed observations conducted by
our group. An important data set is that provided by the Pilot Survey
conducted by the EIS project using the wide-field imager (WFI) on the
MPG/ESO 2.2m telescope. This survey was conducted in $B$ and $V$
passbands and covers over 14 square degrees of the region originally
surveyed by the NTT $I$-band survey. In addition, observations have been
conducted in the $R$-band using the Danish 1.5m telescope at La Silla
and in the infrared $J$ and $K_s$ passbands using SOFI at the NTT.
Currently, the available data for the candidate EIS clusters consists
of: 310 $\gsim3\sigma$ cluster candidates drawn from the $I$-band
survey; 163 candidates with $BVI$; 75 candidates with $BVRI$; and over
35 candidates with $BVRIJK_s$.  Different combination of these data have
been used to identify early-type red-sequences, to estimate photometric
redshifts, and to select targets for spectroscopic observations of
cluster candidates at different redshifts.

\begin{figure*} 
\centering
\includegraphics[width=0.5\textwidth]{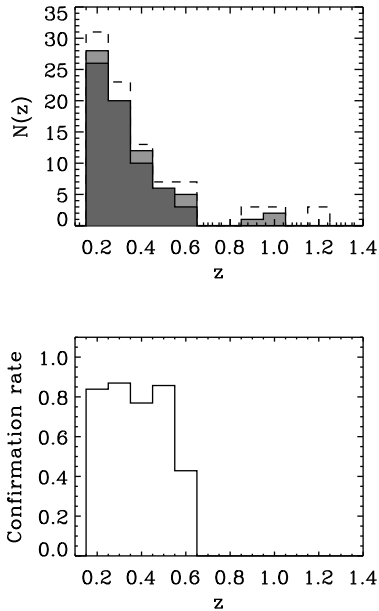}
\caption[]{Redshift distribution (upper panel) of $4
\sigma$ candidates identified by the matched-filter technique, those
identified in the $V$ and $I$ catalogs (light gray), and those
identified by the detection of a red-sequence (dark gray) as described
in the text. In the lower panel the fraction of candidates with
detected concentration in a $(V-I)$ color bin, indicating the
clustering of early-type galaxies, is shown.}
\label{fig:yield} 
\end{figure*}

The matched-filter algorithm has been applied to a sub-sample of the
Pilot Survey $V$-band data and the identified clusters have been
compared to the $I$-band detections~\cite{olsenthesis}. By considering
the number of successful matches between these two samples we estimate
the fraction of real systems to be $\gsim$60\% out to $z\sim0.5$,
increasing to $\gsim$80\% at lower redshifts.  Another way of
confirming the cluster candidates is to combine the $VI$ data to
explore the color properties of the galaxies. For instance,
low-redshift candidates exhibit a well-defined red-sequence in the
$(V-I)-I$ color-magnitude diagram, a clear indication of a
concentration of red-objects.  For more distant clusters, however, the
detection of a red-sequence requires a search for significant
concentrations in the projected distribution of galaxies split in
color bins.  These results are summarized in Figure~\ref{fig:yield}
which shows the redshift distribution of $4
\sigma$ candidates identified by the matched-filter technique, those
identified in both the $V$ and $I$ catalogs, and those identified by
the detection of a red-sequence. Note that both methods yield
consistent results. The usefulness of $VI$ data is limited to systems
with $z\lsim0.5$.  For high-redshift candidates a similar approach has
been used for candidates having $IJK_s$ data.  Out of 11 candidates
analyzed so far, nine are consistent with them being at $z>0.6$ when
compared with the colors of early-type galaxies in clusters
spectroscopically confirmed at high-redshifts.

For low-redshift clusters ($z\lsim0.4$) the matched-filter method has
been complemented by an independent search technique based on the
color information.  We use the available $BVI$ data to classify
early-type galaxies in different redshift intervals. This was done by
adopting a model for galaxy evolution and by selecting galaxies in the
region of the $(B-V)-(V-I)$ color space likely to be occupied by
early-type galaxies in a given redshift interval. The projected
distribution of the resulting sample of early-type galaxies was then
used to identify significant concentrations. The main advantages of
the method are that it uses the color properties of galaxies,
minimizes the effects of foreground/background contamination and
candidate cluster members can be identified. While largely independent
of the matched-filter, since no assumptions are made about the cluster
profile and luminosity function, the majority of the candidate
clusters are in common with those identified by the matched-filter.

\begin{figure*}
\centering
\includegraphics[width=0.5\textwidth]{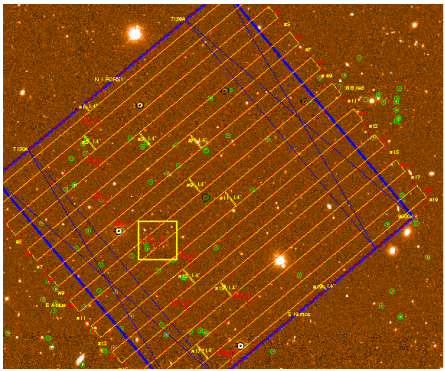}
\caption[]{Example of a FORS mask used in the observation of a
candidate
cluster at $z\sim0.8$. The square indicates the location and scale of
the main galaxy concentration. }
\label{fig:fors}
\end{figure*}

Finally, for candidates with more than four passbands ($BVIJK_s$ and
$BVRIJK_s$) photometric redshifts have been estimated. The galaxy sample
was then split into photometric redshift bins and a search was conducted
to confirm the existence of a concentration at the expected location of
the cluster. Galaxies in the redshift interval of the candidate cluster
were selected and have been used as targets in follow-up spectroscopic
observations of the high-redshift clusters, as discussed below. This is
essential in order to increase the efficiency of the spectroscopic
observations, especially in the case of spectrographs such as FORS1 with
a large field of view compared to the size of the core of the cluster. 
Figure~\ref{fig:fors} shows one of the masks utilized in the observation
of a distant cluster. The main galaxy concentration is indicated by the
square shown in the figure. As can be seen, for distant clusters many
different masks must be used to properly sampled the cluster core.
Furthermore, in order to sample the outskirts of the cluster it must be
possible to select galaxies with different SEDs with redshifts
comparable to that estimated for the candidate cluster in order to
minimize foreground/background contamination.

Recently we have also used FORS to obtain a deep $R$-band image of one
high-redshift candidate cluster under excellent seeing conditions. A
preliminary analysis of the data suggests a significant weak lensing
signal nearly coincident with the concentration of galaxies associated
with the matched-filter detection. If confirmed, this would be the
first detection of weak lensing from an optically-selected cluster and
a strong evidence that at least some of the EIS candidate clusters are
indeed associated with mass concentrations.

\section {Spectroscopy}

In addition to photometric observations there are several ongoing
programs, targeting different redshift intervals, to confirm the EIS
clusters spectroscopically. For systems with $z\lsim0.4$, the first
attempt, carried out using the 2dF spectrograph~\cite{colless},
provided encouraging results.  Despite poor sampling, poor constraints
in the selection of the individual targets, and the fact that the
observations were conducted under less than ideal conditions, some 22
clusters were tentatively confirmed, representing a yield of
$\gsim60\%$. We believe that by using the multi-color data now
available it will be possible to greatly improve the sampling of
galaxies in clusters leading, perhaps, to enough measurements to allow
estimates of the velocity dispersion to be made.

Observations of clusters selected with matched-filter estimated
redshifts in the range 0.5-0.7 have also been conducted using the 3.6m
telescope at La~Silla.  Preliminary results have been reported
elsewhere~\cite{ramella}. At the time of writing a total of 15
clusters have been observed and eight, out of the 10 reduced, have
been assigned redshifts in the range 0.44-0.67. While there are enough
measurements per cluster to allow their detection in redshift space,
the number of galaxies with measured redshifts is still too small to
estimate the velocity dispersion of these systems.

\begin{figure*}
\centering
\includegraphics[width=0.5\textwidth]{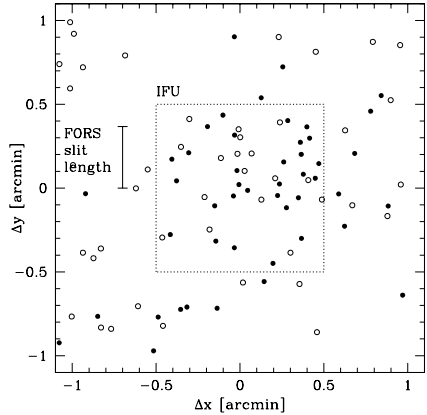}
\caption[]{Example of a z=0.8 candidate cluster from our sample.
The projected distribution of galaxies in the central region of the
cluster is presented. The filled circles correspond to galaxies
brighter than $I_{AB}=22.5$ and the open circles those brighter than
$I_{AB}=23.5$. Also shown is the size of a FORS slit compared to the
IFU field of view. }
\label{fig:ifu}
\end{figure*}

Recently, spectroscopic observations have also been carried out using
FORS1 at the VLT. A total of seven candidate systems have already been
observed. Preliminary results are encouraging with the four candidates
partially analyzed showing a large number of coincident redshifts in
good agreement with those estimated by the matched-filter and
photometric redshift analyses. However, Figure~\ref{fig:fors} shows
that while FORS is adequate for studying the outer parts and the
environment of a confirmed cluster, confirmation and measurement of
the cluster velocity dispersion for a large number of distant clusters
cannot be efficiently done with it. This will only be possible using
the integral field unit (IFU) of the VIMOS spectrograph.
Figure~\ref{fig:ifu} shows the core of a cluster at $z\sim~0.8$
compared to the height of a FORS slit. In the central 1~arcmin$^{2}$,
$\sim 90\%$ of the galaxies brighter than $I_{AB}=22.5$ (filled
circles) have been found to be potential cluster members using the
photometric redshift technique. Typical galaxy densities in the core
of a $z\gsim~0.8$ cluster candidate are: $\sim$ 25 galaxies
arcmin$^{-2}$ at $I_{AB}=22.5$ and $\sim$ 45 galaxies arcmin$^{-2}$ at
$I_{AB}=23.5$. Therefore, using the IFU, with a field of view of $54"
\times 54"$, it will be possible to observe over 30 galaxies in a
single pointing, thereby enabling us to: 1) densely sample the
candidate cluster core, minimizing the effects of interlopers; 2)
probe sub-L$_*$ galaxies; and 3) have a complete spectroscopic sample
down to $I_{AB}=23.5$. The IFU thus provide a considerable improvement
for measurements of the velocity dispersion and for estimating the
virial mass of high-z clusters.

\section{Summary}

Using multi-color data and preliminary results from ongoing
spectroscopic observations we find that a significant number of EIS
cluster candidates are real physical associations as determined
indirectly from the presence of a concentration of early-type galaxies
or directly from measured redshifts. Though based on an admittedly small
sample in some redshift ranges, the yield from the original cluster
catalog is better than 60\%. Some 80 clusters with $z\lsim0.6$ and 7
with $z\gsim0.8$ have already been confirmed. This number should rapidly
increase as new data are reduced.

Using the above estimate for the yield we expect that it will be
possible to construct a sample of confirmed clusters consisting of
some 150 clusters with $z\lsim0.6$ and some 80 with $z\gsim0.6$. The
upper limit in the cluster redshift for the sample is still poorly
determined and must await new spectroscopic observations with the
integral field unit of VIMOS. These observations are also essential in
order to efficiently study the dynamics of high-redshift clusters.

\clearpage
\addcontentsline{toc}{section}{Index}
\flushbottom
\printindex

\end{document}